# Crystallization of Crystalline/Crystalline Symmetric Polymer Blends Studied by Dynamic Monte Carlo Simulations

Ashok Kumar Dasmahapatra[*]

Department of Chemical Engineering, Indian Institute of Technology Guwahati, Guwahati – 781039, Assam, India

---
[*] Corresponding author: Phone: +91-361-2582273, Fax: +91-361-2582291, Electronic mail: akdm@iitg.ernet.in

**ABSTRACT**

Blending of two or more pure polymers is an effective way to produce composites with tunable properties.    In this paper, we report dynamic Monte Carlo simulation results on the crystallization of crystalline/crystalline (A/B) symmetric binary polymer blend, wherein the melting temperature of A-polymer is higher than B-polymer.    Crystallization of A-polymer precedes the crystallization of B-polymer upon cooling from a homogeneous melt.    The morphological development is controlled by the interplay between crystallization and macrophase separation between two polymers.    With increasing segregation strength, mean square radius of gyration shows a decreasing trend, reflecting the increased repulsive interaction between A- and B-polymers.    As a consequence, a large number of smaller size crystals form with lesser crystallinity.    Isothermal crystallization reveals that the transition pathways strongly depend on segregation strength.    We also observe a path-dependent crystallization behavior in isothermal crystallization: two-step (sequential) isothermal crystallization yields superior crystalline structure in both A- and B-polymers than one-step (coincident) crystallization.

## I. INTRODUCTION

Multi-component polymeric systems[1,2] (viz., mixture of two or more polymers) paves the pathway to prepare tailor made polymer composites for a wide range of modern applications, such as in polymer-based light emitting diodes [3] (LED), medical appliances[4] and nanotechnology.[5] The extent of property enhancement strongly depends on the mutual miscibility among the components. The chemical dissimilarity leads to the formation of macrophase separated domains[6-8] influencing morphology of the resultant composites. For a miscible blend, a single glass transition temperature (Tg) is observed,[9-17] however for immiscible polymer blends, two distinct Tgs for individual component are observed.[18]

Decades of research has been devoted to understand phase behavior, miscibility pattern and crystallization characteristics of binary blend with one crystallizable component. The interplay between macrophase separation, and crystallization of the crystallizable component would dictate the crystallization behavior and final morphology. In most of the cases it has been found that the amorphous component is excluded from the crystalline domains and resides between bundles of lamellar stacks[19,20] which in turn retard the overall crystallization[21-24] with the formation of a diverse range of morphological pattern.[25]

Although binary polymer blend with one crystalline component has been extensively studied, the crystallization and morphological development of crystalline/crystalline binary blend poses promising opportunity for developing novel materials. The richness of the phase behavior directly depends on the mutual miscibility and individual crystallizability of the constituting components. Therefore, the resultant morphology of the composite would be a complex one. The difference in melting point plays a crucial role in determining crystallization sequence. In most of the cases, the high melting component (HMC) crystallizes first and the low melting component (LMC) is excluded from the crystalline domains of HMC. On further cooling, LMC crystallizes and fill the space. LMC crystals (spherulites or lamella) may grow in the inter-spherulites space of HMC. However, mixed crystals (spherulites or lamella) are usually not possible due to crystallographic constraints. The crystallization of LMC is usually retarded due to the presence of HMC crystals all over the space, and hence crystallinity of LMC is less than HMC.[10] For example, in Poly(3-hydroxy butyrate)/ Polye(thylene Oxide) (PHB/PEO) blend, PHB crystallizes first followed by the crystallization of PEO.[9,26,27] Crystallization of PEO is hindered in the presence of already crystalline domains of PHB, leading to less crystalline materials. Presence of PEO decreases the melting point of PHB,[28] due to the formation of a miscible pair, showing a single $T_g$. Similar depression of $T_g$ has also been reported for PHB/ Poly(Butylene Succinate) (PBS) blend.[29] Melt miscibility is also dictated by the molecular weight of the constituting components. For example, low molecular weight Poly(L-Lactic Acid) (PLLA) is miscible with PHB, however high molecular weight PLLA is not, influencing mechanical properties[30-32] of the composite. Blends with closely spaced melting points lead to simultaneous crystallization[33-35] with the formation of interpenetrating spherulites.[36]

Apart from the melting point difference, the crystallization behavior is also dictated by the mutual immiscibility, which is governed by the Flory's $\chi$ parameter. The magnitude of $\chi$ is governed by the combination of constituent components of a blend. Consequently, the extent of macrophase separation and resultant semi-crystalline morphology changes, which can successfully be exploited to tune the properties of the composites. For example, in the presence of PEO in PHB/PEO blend, the melting point of PHB decreases. However, crystallization of PEO is hindered by PHB.[9,26,27] On the other hand, PHB/PLLA blend produces interpenetrating spherulites during crystallization[30-32] with a faster shperulite growth for PHB compared to PLLA;[30] PBS in PBS/PHB blend decreases the melting and glass transition temperature of PHB.[29] Poly(carbonate) in Poly(carbonate)/ Poly(ε-caprolactone) (PC/PCL) blend restricts the crystallization of PCL and a composition dependent degradation behavior of PCL is observed;[37] presence of PBS decreases crystallization temperature and retards the crystallization of PCL in PBS/PCL blend;[38] PEO does not affect the crystallization of PCL, however, the crystallization of PEO is retarded by PCL in PCL/PEO blend.[39]

In most of the cases crystallization of LMC is retarded in the presence of already crystalline domains of HMC. However, it has also been observed that both the component may crystallize simultaneously (viz., coincident crystallization) where their melting points are relatively close to each other, and one component accelerates the crystallization of the other. For example, in PLLA/PBS blend, PBS accelerates the crystallization of PLLA, which is attributed to the fact that the PBS acts as a heterogeneous nucleating agent for the crystallization of PLLA.[40] Addition of Poly(Ethylene Succinate) (PES[41]) and low molecular weight PHB[11] increases the rate of crystallization of PLLA in PLLA/PES and PLLA/PHB blends respectively.

In this paper, we describe simulation results on crystallization of binary polymer blend with equal composition of two crystallizable components to explore how the crystallization of one component influences the crystallization of other. Our results show a monotonically decreasing trend in crystallinity with the formation of smaller and thinner crystallites on increasing the magnitude of segregation strength. Isothermal experiments reveal a strong dependency of the segregation strength on transition pathway.

## II. MODEL AND SIMULATION TECHNIQUE

We employ a lattice based coarse-grained model to simulate crystallization of polymer blends. To simulate a bulk polymer system, we put 480 chains in a simple cubic lattice of size 32x32x32, so that the occupation density is as high 0.9375, representing a melt system. Initially, we put polymer chains along lattice grid keeping the connectivity between units intact. Each polymer chain consists of 64 repeat units. For a symmetric binary blend, we use 240 polymer chains of either type (viz., A- and B-polymer). We apply a set of microrelaxation moves to generate a homogeneous melt. We employ single site bond fluctuation algorithm along with periodic boundary conditions to move chain molecules along the lattice grids. The coordination number of our lattice is 26 (6 along the lattice axis, 12 along the face diagonals and 8 along the body diagonal). Thus, bond length can be 1 (along the axis), $\sqrt{2}$ (along the face diagonal) or $\sqrt{3}$ lattice units (along the body diagonal). We select a vacant site randomly from the available vacant sites and search for a nearest site occupied by either A- or B-type units. Once an occupied site is found, an appropriate microrelaxation move is selected according to their position along the chain. To give further details, if the selected unit is at the end of a chain, then end bond rotation and slithering diffusion is selected with equal probability. On the other hand, if the unit is non-terminal, a single site bond fluctuation is implemented. [42]

During the movement we have strictly implemented excluded volume effect – one lattice site is occupied by one unit (either A- or B-type), and no bond crosses each other. Once this self-avoiding criterion is satisfied, we calculate the change in energy associated with the movement. The crystallization driving force is modeled as an attractive interaction between neighboring parallel bonds and collinear bonds within A- or B- type units and given by $U_p$ and $U_c$ respectively. The interaction between A- and B-type units is modeled as a repulsive interaction to represent their mutual immiscibility. The energy penalty to create A-B contact is modeled by $U_{AB}$. The change in energy for a Monte Carlo (MC) move is given by:

$$\Delta E = -\left(\Delta N_p U_p + \Delta N_c U_c\right)_A - \left(\Delta N_p U_p + \Delta N_c U_c\right)_B + \Delta N_{AB} U_{AB}$$

Where, $\Delta N_p$ and $\Delta N_c$ represents the net change in the number of parallel and collinear bond respectively for the A and B polymer, and $\Delta N_{AB}$ represents the change in the number of contacts between A- and B-units.

We model the binary blend such that both the polymers (A and B) are crystallizable while the melting temperature of the B-polymer is lower than that of A-polymer. Thus, low-melting B-polymer is less facile to crystallize upon cooling from a high temperature melt. Therefore, the crystallization driving force of B-polymer is relatively less in comparison to that of A-polymer. To implement this, we use $U_{pB} = \lambda_m U_{pA}$ and $U_{cB} = \lambda_m U_{cA}$ for the parallel and collinear bond interaction energy respectively. We set $\lambda_m = 0.75$ (<1) to represent that B-polymer has less driving force for crystallization compared to A-polymer. Further, we assume that $U_p = U_c$, for the coarse grained interactions used in our simulation. The interaction between A- and B-polymer, $U_{AB}$ is calculated as $\lambda U_p$, where $\lambda$ represents the segregation strength (viz., A-B demixing energy) relative to the parallel bond interaction energy, and equivalent to Flory's $\chi$ parameter. In our work $\lambda \geq 0$ and higher $\lambda$ implies stronger segregation strength between the polymers. In terms of Flory's $\chi$ parameter, segregation strength is calculated as $\chi N$, which may be correlated to $(q-2) \times U_{AB} \times N$ in our system,[43-45] where $q$ is coordination number and $N$ is the degree of polymerization. All the energies are normalized by $k_B T$, where, $k_B$ is the Boltzmann constant and $T$ is temperature in Kelvin; thus, $U_p \sim 1/T$. Now the change in energy per MC move is modified as follows:

$$\Delta E = \left[\left(\Delta N_p + \Delta N_c\right)_A + \lambda_m\left(\Delta N_p + \Delta N_c\right)_B + \lambda \Delta N_{AB}\right] U_p \tag{1}$$

We employ the Metropolis sampling scheme with periodic boundary conditions to sample new conformation. The probability of an MC move is given by $\exp(-\Delta E)$. We accept new conformation if $\exp(-\Delta E) \geq r$, where $r$ is a random number in the range (0, 1), generated by using the random number generator, MT19937. [46] We simulate the crystallization of binary blend by varying $U_p$ from 0 (viz., at infinite temperature, athermal state) to 0.6 with a step size 0.02, to represent a step-cooling from a high temperature melt. To equilibrate the system we calculate mean square radius of gyration, $\left\langle R_g^2 \right\rangle$, as a function of Monte Carlo Steps (MCS). We did not observe an appreciable change in the value of $\left\langle R_g^2 \right\rangle$ beyond 5000 MCS and it is considered as the equilibration time. We calculate thermodynamics and structural parameters averaged over subsequent 5000 MCS.

To monitor transition from a disordered melt to an ordered crystalline phase, we calculate fractional crystallinity, $X$ of A-polymer, B-polymer and overall, and specific heat ($C_v$) as a function of $U_p$. We define crystallinity as the ratio of crystalline bonds to the total number of bonds present in the system. A bond is defined as crystalline if it is surrounded by more than 5 nearest non-bonded parallel bonds.[44,45] Specific heat is calculated as equilibrium specific heat from the total energy fluctuations (for all the A- and B-type units in the simulation box), similar to that of Dasmahapatra et al.[47] To locate the macrophase separation point, we calculate $C_v$ for A-B pair ($C_{v\_AB}$) based on the de-mixing energy between A and B-polymers. [44,45] At the macrophase separation point, $C_{v\_AB}$ shows a peak; peak value of $U_p$ decreases with increasing the value of $\lambda$. During macrophase separation, individual polymer forms their respective domains, and the resulting transition may be considered as a disordered to ordered transition, as a result of which polymer chain shrinks, which is reflected by a decreasing trend of $\left\langle R_g^2 \right\rangle$ with increasing $U_p$ (viz., decreasing temperature). We also calculate average crystallite size and lamellar thickness. A crystallite of size $S$ is defined as a small microscopic aggregate having $S$ number of crystalline bonds with same orientation. We calculate the average crystallites size as a function of $U_p$. We express lamellar thickness as the average number of monomer units present in a given crystallite (in the direction of crystal thickness), and average thickness is calculated over all crystallites present in the system.

## III. RESULTS AND DISCUSSION

We begin with describing the details of non-isothermal crystallization of 1:1 binary polymer blend containing total 480 polymer chains (240 chains each for A- and B-polymer). Each chain (both A- and B-polymer) consists of 64 repeat units. Subsequently, we discuss isothermal crystallization (one- and two-step) with the detail analysis on development of crystal morphology.

### III.A. Monitoring Phase Transition

We equilibrate the sample system at $U_p = 0$ and gradually cool to $U_p = 0.6$ with a step size 0.02. The step-wise cooling corresponds to a non-isothermal crystallization. Polymer chain segments arrange in parallel orientation as we cool the system by increasing the value of $U_p$. At $U_p = 0$, we generate a homogeneous melt, where A- and B-units are evenly dispersed (Figure 1). Increasing $U_p$ leads to the transition from a disordered melt to an ordered crystalline state. We monitor the crystallization of binary melt by following the change in equilibrium specific heat,[48] calculated from the energy fluctuation as a function of $U_p$. $C_v$ vs. $U_p$ shows a peak at a certain value of $U_p$, which is regarded as the transition point from melt to crystal in tune with the experimental observation.[49] Figure 2a shows the variation in $C_v$ with $U_p$ for a series of segregation strength: $\lambda = 0$ to 6. Since A- and B-polymers have different melting point (the way it has been modeled), we see two transition points for A- and B-polymer respectively.[50-53] A-polymer crystallizes first followed by the crystallization of B-polymer, since the crystallization driving force for A-polymer is higher than that of B-polymer. Transition point (in terms of $U_p$) remains same for $\lambda = 0$, 1, 2 and 3 for both the polymers. However, for $\lambda = 4$, 5 and 6, the transition point shifts towards higher value of $U_p$ (viz, lower temperature) for both the polymers (Figure 2b). As the segregation strength increases, de-mixing energy between two polymers increases affecting the transition of both the polymers. Higher cooling (higher value of $U_p$) is needed to overcome the resistance to facilitate crystallization. Our simulation results are in close agreement with the experimental observation on blend of PHB/PLLA (50/50), where melting point of PHB decreases on addition of PLLA.[30] Usually, the depression of melting point of fist component is observed on addition of second component[28,30]; however, depression of melting point for both the component is also observed, depending on the combination of A- and B-polymers.[10,35,38,54,55]

We have calculated $\langle R_g^2 \rangle$ for both the polymers as a function of $U_p$ to understand demixing behavior of A- and B-polymer. Figure 3a shows that with increasing $U_p$, the value of $\langle R_g^2 \rangle$ suddenly decreases, indicating the formation of individual domain of A- and B-polymer as a result of the repulsive interaction between A and B (viz., onset of macrophase separation). Beyond $U_p \sim 0.3$, there is practically no change in the value of $\langle R_g^2 \rangle$ indicating the formation of crystalline domains of A-polymer, and therefore, no further change is anticipated. The saturation value of $\langle R_g^2 \rangle$, at the end of crystallization (viz., at $U_p = 0.6$), shows a marginal variation with $\lambda$ (Figure 3b). We have also calculated $C_{v\_AB}$ to correlate with $\langle R_g^2 \rangle$ trend. With increasing $\lambda$, the macrophase separation happens at a relatively higher temperature (viz., lower value of $U_p$) (Figure 3b inset), signifying that higher segregation strength promotes de-mixing and, it happens prior to the crystallization. Once the macrophase separation occurs, the individual component (A and B) crystallize within their respective domain, and with a suppressed crystallization as seen in $C_v$ trend (Figure 2). Figure 4 shows snapshots of macrophase separated melt for $U_p = 0.02$ and 0.1 for $\lambda = 1$ and 3.

## III.B. Development of Crystallinity during Non-isothermal Cooling

We study the development of crystallinity during non-isothermal crystallization by calculating crystallinity of A-polymer ($X_A$), B-polymer ($X_B$) and overall ($X$) as a function of $U_p$, for a series of $\lambda$. Figure 5a shows an abrupt increase in $X_A$ at $U_p \sim 0.3$ for all the values of $\lambda$ and reaches at a saturation crystallinity at $U_p \sim 0.5$. We observe a decrease in saturation crystallinity with increasing segregation strength, $\lambda$: $X_A^{sat}$ decreases from 0.75 at $\lambda = 0$ to 0.51 at $\lambda = 6$. With increasing segregation strength, the repulsive interaction dominates between two polymers, resulting slowing down chain mobility and hence inhibits crystallization and development of crystallinity. We observe a similar trend in case of B-polymer, $X_B$ vs. $U_p$, where the sudden increase in $X_B$ happens at a relatively higher $U_p$ (viz., lower temperature), $U_p \sim 0.4$ (Figure 5b), since B-polymer is low melting polymer compared to A-polymer. At $U_p \sim 0.55$, $X_B$ almost reaches its saturation value. As shown by A-polymer, B-polymer also exhibits a decreasing trend in saturation crystallinity with increasing value of $\lambda$: $X_B^{sat}$ decreases from 0.76 at $\lambda = 0$ to 0.35 at $\lambda = 6$. However, the saturation crystallinity is higher for A-polymer than B-polymer except for all $\lambda$ except $\lambda = 0$. Since A-polymer is more crystallizable, it crystallizes first within its own domain. With increasing the value of $\lambda$, crystallinity decreases for the both the polymer due to the decrease in chain mobility. We also examine the trend in overall crystallinity, $X$ (combined crystallinity of A- and B-polymers) as a function of $U_p$ for all $\lambda$ (Figure 5c). The trend in $X$ vs $U_p$ clearly demonstrates the two regimes of crystallization for two crystallizable polymers. For all values of $\lambda$, $X$ increases abruptly at $U_p \sim 0.3$ (where A-polymer starts to crystallize) reaches its saturation crystallinity at $U_p \sim 0.55$ after a slow increase (where A-polymer almost reaches the saturated crystallinity and B-polymer begins to crystallize) at $U_p \sim 0.35$. The overall saturation crystallinity also shows a monotonically decreasing trend with $\lambda$: $X^* = 0.75, 0.74, 0.70, 0.65, 0.59, 0.52$ and $0.43$ for $\lambda = 0, 1, 2, 3, 4, 5$ and 6 respectively. With increasing segregation strength, A- and B-polymer becomes more immiscible with each other and forms respective domains via macrophase separation. The repulsive interaction mainly operates on the interface between A- and B-domains, leading to decreasing the chain mobility of the segments on the surface. This decrease in chain mobility on the surface affects the mobility of the segments inside the domain. As the crystallization happens within these small domains, and the overall energy depends on the both A-B interaction and crystallization driving force (see Equation 1), the development of crystallinity of both the components is restricted. Our simulation results are in close agreement with the experimental observation, where high melting component crystallizes first and suppresses the crystallization of the low melting component.[12,56] Gradual heating from $U_p = 0.6$ to $U_p = 0$ with a heating step 0.02 exhibits a hysteresis on melting (Figure 6). This pattern is quite similar for non-equilibrium, first order phase transition (viz., crystallization-melting) in bulk polymers.

### III.C. Structural Analysis

We estimate the change in average crystallites size and lamellar thickness as we crystallize the system. Crystallites size shows a wider distribution compared to lamellar thickness. The magnitude of crystallites size is much higher than the lamellar thickness of the crystallites, indicating the formation of two-dimensional crystals in the system. We observe that the average crystallites size of A-polymer, $\langle S_A \rangle$ and B-polymer, $\langle S_B \rangle$ increases with increasing $U_p$ and saturates at $U_p \sim 0.55$ (Figure 7a, b), similar to that of $X_A$ and $X_B$ respectively (Figure 5a, b). Abruptness in values of $\langle S_A \rangle$ and $\langle S_B \rangle$ has been observed for lower value of $\lambda$ ($\lambda \leq 4$). For higher values of $\lambda$ (viz., $\lambda = 5$ and 6) a marginal increase in $\langle S_A \rangle$ and $\langle S_B \rangle$ has been observed. High segregation strength restricts the development of crystalline domain and results in the formation of large number of smaller size crystallites, which has also been reflected in the crystallinity trend (Figure 5).

The trend of lamellar thickness of A- and B-polymer is identical to that of $\langle S_A \rangle$ and $\langle S_B \rangle$. The saturation lamellar thickness (viz., at $U_p = 0.6$) for A-polymer (Figure 8a) and B-polymer (Figure 8b) shows a decreasing trend with $\lambda$, signifying the formation of thinner crystals against the usual tendency of crystal thickening in pure homopolymers. Higher segregation strength (viz., $\lambda > 4$) results in the formation of smaller size domains of A- and B-polymer, which upon crystallization produces smaller and thinner crystals. For lower value of $\lambda$ (viz., weak segregation strength), $\langle l_A \rangle$ and $\langle l_B \rangle$ are close to each other, however for higher value of $\lambda$, $\langle l_A \rangle$ is always higher than $\langle l_B \rangle$. At lower value of $\lambda$, the crystal thickening is not affected but at higher value of $\lambda$, the effect is quite prominent –due to the formation of smaller size domains (act as confinement) crystal thickening is restricted. Hence, we observe a decreasing in lamellar thickness with increasing $\lambda$. Figure 9 displays the snapshots from our simulation for $\lambda = 1$, at $U_p$ = 0.28 and $U_p$ = 0.6. At $U_p$ = 0.28, A-polymer is almost crystallized forming lamellar morphology (blue and magenta lines represent crystalline segment of A- and B-polymers respectively, yellow lines represents non-crystalline segments of both the polymers). When $U_p$ is increased to 0.6, B-polymer is crystallized (yellow coloured lines are converted to magenta coloured lines), keeping the morphological pattern intact. We present snapshots from our simulation at $U_p$ = 0.6 in Figure 10 to compare the development of crystalline morphology as a function of $\lambda$ ($\lambda$ = 0, 1, 2, 3, 4, 5 and 6). These snapshots clearly show the decrease in crystallinity with smaller crystallite size.

We have presented our simulation results by varying the segregation strength, $\lambda$ from 0 to 6. Variation of $\lambda$ (viz., variation of segregation strength) can be realised by keeping one polymer fixed and varying the other one to prepare a series of binary blends. For a binary blend, properties of A-polymer can successfully be tuned by varying different B-polymers. For example, PC in PCL/PC blend,[37] increases the degradation temperature of PCL with a decrease in melting point, and reduction in crystallinity; PBS in PCL/PBS blend,[38] retards the crystallization of PCL; whereas, PEO in PCL/PEO blend,[39] does not affect the crystallization of PCL, however, the crystallization of PEO is affected by PCL. Similarly, PEO in PLLA/PEO blend [57] increases the toughness of PLLA; PHB in PLLA/PHB blend [30] increases the melting point of PLLA. Therefore, by changing B-polymer (viz., PC, PBS or PEO) in PCL/B binary blend, the properties of PCL and the blend can suitably be tuned for specific applications.

### III.D. Isothermal Crystallization

To carry out isothermal crystallization, we first equilibrate the sample system at $U_p$ = 0 (athermal state). We quench the sample system to $U_p$ = 0.6, annealed for $10^5$ Monte Carlo steps (MCS) and measure $X_A$ (for A-polymer), $X_B$ (for B-polymer) and $X_c$ (overall) as a function of MCS. Figure 11a represents the evolution of crystallinity with MCS for all the $\lambda$ values investigated. We also calculate scaled crystallinity, $X_c^* = \left( X_c - X_c^i \right) \Big/ \left( X_c^f - X_c^i \right)$, ranges from 0 to 1.0 as a function of MCS for all values of $\lambda$ (Figure 11b). $X_c^i$ represents the crystallinity at the beginning of the isothermal experiment, and $X_c^f$ represents the crystallinity at the end of isothermal annealing (viz., $X_c$ at the end of $10^5$ $MCS$). The trend in overall crystallinity as a function of $\lambda$ reveals that the kinetic pathway for crystallization of two crystallisable components is a strong function of their mutual immiscibility, governed by the value of $\lambda$. Figure 12a and 12b compare the early stage (up to 1000 MCS) of crystallization of both the polymers at $\lambda = 0$ and 2 respectively. For $\lambda = 0$, both the polymers show almost same crystallinity during that period till MCS ~ 10, beyond that they follow different pathway with A-polymer producing higher crystallinity than B-polymer, since A-polymer has higher crystallization driving force than B-polymer. For $\lambda = 2$, A- and B-polymers follows different pathway even early (<10 MCS) with A-polymer producing higher crystallinity than B-polymer. At $\lambda = 2$, macrophase separation takes place at a very early stage leading to different time evolution trajectory for A- and B-polymer. The overall saturation crystallinity (viz., crystallinity at the end of $10^5$ MCS), for $\lambda = 0$ is close to that of non-isothermal crystallization. However, the value decreases monotonically with increasing $\lambda$. For example, saturation crystallinity in isothermal crystallizaiton for $\lambda = 0, 1, 2, 3, 4, 5$ and 6 are 0.71, 0.65, 0.53, 0.28, 0.14, 0.093 and 0.066 respectively (Figure 11a). The corresponding values during non-isothermal crystallization are: 0.75, 0.74, 0.70, 0.65, 0.59, 0.52 and 0.43 for $\lambda = 0, 1, 2, 3, 4, 5$ and 6 respectively (Figure 5c). From the above analysis, it is evident that at high segregation strength, isothermal crystallization produces low crystalline material than the non-isothermal crystallization, due to the lack of thermodynamic stability needed for equilibrium chain folding.

In our present binary polymer melt, we model A- and B- polymer with different crystallizability and we expect that the quench depth may influence crystallization and morphological development. We carried out isothermal experiment in two steps as follows: in the first step we equilibrate at $U_p = 0$ and quench to $U_p = 0.28$, annealed for $10^5$ MCS; and in the second step, we quench to $U_p = 0.6$ from $U_p = 0.28$. At $U_p = 0.28$ (temperature above the melting point of B-polymer), A-polymer crystallizes, while B-polymers are still in a molten state (Figure 13a). Upon quenching to $U_p = 0.6$ (from $U_p = 0.28$), B-polymer starts to crystallize within the domains created during macrophase separation followed by A-polymer crystallization (Figure 13b). We observe that there is a significant enhancement of crystallinity of both A- and B-polymer (Figure 13c) in comparison with one-step quenching as described above (see Table 1). We interpret this enhancement in crystallinity as follows. When we quench the sample system to $U_p = 0.28$, temperature below the melting point of A-polymer, but above the melting point of B-polymer, A-polymer crystallizes within the macrophase separated A-domains and without much hindrance of B-polymer (since B-polymer is in macrophase separated melt state). In the next stage, when quench to $U_p = 0.6$, below the melting point of B-polymer (also A-polymer), crystallization of B-polymer (happens within the B-domains) does not experience hindrance from A-polymer, since A-polymer already crystallized within its own domains. Thus, the mode of crystallization in case of two-step cooling appears to be sequential crystallization, as evident by the recent experimental observation on PLLA/PEO [57] and PEO/PES blend.[58] In PLLA/PEO blend, crystallinity and spherulite size decreases with decreasing crystallization temperature. In PEO/PES blend ($T_m^{PEO} \approx 59$ °C, $T_m^{PES} \approx 101$ °C) one-step (cooling at 40 C) and two–step (first cooling at 70 °C followed by cooling at 40 °C) yields different morphological pattern. In one-step isothermal crystallization, when we quench to $U_p = 0.6$, the temperature is well below the melting point of both A- and B-polymer, and hence the driving force for the crystallization of both the polymers is adequate. In our model, the crystallization driving force for A-polymer is higher than B-polymer; as a result, A-polymer possesses higher crystallinity than B-polymer as the crystallization progresses with Monte Carlo steps. Thus the mode of crystallization is referred to as coincident crystallization, where both the polymers compete with each other to crystallize simultaneously. This observation is in close agreement with the experimental observations.[58,59] In one-step isothermal crystallization (viz., quenching to $U_p = 0.6$), inter-chain entanglement restricts the development of crystallinity of A-polymer. However, the magnitude of saturation crystallinity of B-polymer in both the experiments is close to each other. The saturation crystallinity of B-polymer is 0.65 and 0.68 (Table 1) for one-step and two-step isothermal crystallization, respectively. Thus, two-step quench produces better crystalline structure for A-polymer over B-polymer. Analysis based on lamellar thickness reveals that the lamellar thickness of A-polymer at $U_p = 0.6$ (one-step quenching) is less than that of $U_p = 0.28$, during two-step quench, which is in accord with the Hoffman-Weeks formulation [60] and experimental observation.[61] However, lamellar thickness of B-polymer when compared to $U_p =$

0.6 (one-step cooling and two-step cooling), the values appear identical, as we observed in crystallinity (see Table 2). This is explained as follows. Lamellar thickness is largely dictated by the degree of undercooling $\left( \Delta T = T_m^\alpha - T \right)$. When the system is quenched to $U_p = 0.6$, the effective $\Delta T$ for B-polymer remains same in both one- and two-step isothermal processes. Therefore, the crystal thickening (viz., development of crystal thickness) remains identical in both the cases. Crystal growth along lateral direction continues to produce higher crystallinity.

## IV. CONCLUSIONS

In this paper, we present simulation results on crystallization of a binary polymer blend with equal composition of A- and B-polymers, where both the polymers are crystallizable. In our model A- and B-polymers are miscible at high temperature melt and become immiscible as we cool the sample system from the melt. The segregation strength, $\lambda$ (viz., de-mixing energy) dictates the crystallization behavior and morphological development. With increasing $\lambda$, we observe a gradual decrease in saturation crystallinity with the formation of smaller and thinner crystals of both the polymers. We model A-polymer as high melting polymer, and as a result possess higher crystallinity than B-polymer. As we cool the system from a homogeneous melt, macrophase separation takes place before crystallization. During macrophase separation, A- and B-polymers form their respective domains. With increasing $\lambda$, the domain size decreases due to the increased repulsive interaction between two polymers. As a result, the subsequent crystallization happens within the domains created during macrophse separation producing smaller crystallites. We also observe a decrease in mean square radius of gyration with increasing $\lambda$, indicating the influence of segregation strength on macrophase separation. We monitor the onset of macrophase separation by estimating $C_{v\_AB}$, which shows a peak at a relatively low $U_p$ values compared to that of crystallization. As we increase $\lambda$, the macrophase separation happens at a relatively lower value of $U_p$ (viz., higher temperature). Isothermal crystallization reveals that the crystallization pathway strongly depends on segregation strength as well as mode of cooling (viz., one-step and two-step). Two-step isothermal crystallization follows a sequential crystallization mechanism whereas one-step isothermal crystallization follows a coincident crystallization mechanism. Two-step crystallization process yields a semicrystalline material with higher crystallinity and thicker crystals compared to that of one-step process. In one-step cooling, both the polymers compete with each other for crystallization resulting formation of less crystalline materials. Our simulation results on path dependent crystallization behavior and morphological development of binary polymer blend would be useful to gain insight in tailoring supramolecular properties for potential applications.

## ACKNOWLEDGEMENT

We acknowledge the use of computational facility supported by the SERB, Department of Science and Technology (DST), Government of India (sanction letter no. SR/S3/CE/0069/2010).

**Table captions**

**Table 1: Comparison in fractional crystallinity ( $X$ ) of A- and B-polymer for $\lambda = 1$ during one-step and two-step isothermal crystallization.**

**Table 2: Comparison in average lamellar thickness $\langle l \rangle$ of A- and B-polymer for $\lambda = 1$ during one-step and two-step isothermal crystallization.**

**Figure captions**

**Figure 1. Snapshot of the simulation system representing a homogeneous melt of binary polymer blend composition ( $x_B$ ) = 0.5, at $U_p$ = 0. Blue and magenta line represents segments of A- and B-polymer respectively.**

**Figure 2. (a) Change in specific heat, $C_v$ with $U_p$ for $\lambda$ = 0, 1, 2, 3, 4, 5 and 6. (b) Change in transition point, $U_p^*$ with $\lambda$ for A- and B-polymer. The lines joining the points are meant only as a guide to the eye.**

**Figure 3. (a) Change in mean square radius of gyration, $\langle R_g^2 \rangle$ with $U_p$ for $\lambda$ = 0, 1, 2, 3, 4, 5 and 6. (b) Change in $\langle R_g^2 \rangle$ as a function of $\lambda$ at $U_p$ = 0.6. The inset shows the change in macrophase separation point, $U_p^\#$ (calculated from $C_{v\_AB}$ vs. $U_p$ ) with $\lambda$ . The lines joining the points are meant only as a guide to the eye.**

**Figure 4. Snapshots of macrophase separated melt of (a) $\lambda$ = 1 at $U_p$ = 0.02 (b) $\lambda$ = 1 at $U_p$ = 0.1 and (c) $\lambda$ = 3 at $U_p$ = 0.02 (d) $\lambda$ = 3 at $U_p$ = 0.1 . Blue and magenta line represents segments of A- and B-polymer respectively.**

**Figure 5.** Change in crystallinity of **(a)** A-polymer, **(b)** B-polymer and **(c)** Overall with $U_p$ **(~1/T)** for $\lambda$ **= 0, 1, 2, 3, 4, 5 and 6. The lines joining the points are meant only as a guide to the eye.**

**Figure 6.** Change in fractional crystallinity, $X$ with $U_p$ for $\lambda$ **= 1 during non-isothermal crystallization and melting of binary polymer blend displaying hysteresis. The lines joining the points are meant only as a guide to the eye.**

**Figure 7.** Change in average crystallites size, $\langle S \rangle$ of **(a)** A-polymer and **(b)** B-polymer with $U_p$ at $\lambda$ **= 0, 1, 2, 3, 4, 5 and 6. The lines joining the points are meant only as a guide to the eye.**

**Figure 8.** Change in average lamellar thickness, $\langle l \rangle$ of **(a)** A-polymer **(b)** B-polymer with $U_p$ at $\lambda$ **= 0, 1, 2, 3, 4, 5 and 6. The lines joining the points are meant only as a guide to the eye.**

**Figure 9.** Snapshots of semi-crystalline structure for $\lambda$ **= 1 at (a)** $U_p$ **= 0.28 and (b)** $U_p$ **= 0.6, during non-isothermal crystallization. Blue lines represent crystalline bonds of A-polymer, magenta lines represent crystalline bonds of B-polymer and grey lines represent non-crystalline bonds of both the polymers.**

**Figure 10.** Snapshots of semi-crystalline structure at $U_p$ **= 0.6 for (a)** $\lambda$ **= 0, (b)** $\lambda$ **= 1, (c)** $\lambda$ **= 2, (d)** $\lambda$ **= 3, (e)** $\lambda$ **= 4 and (f)** $\lambda$ **= 5. Blue lines represent crystalline bonds of A-polymer, magenta lines represent crystalline bonds of B-polymer and grey lines represent non-crystalline bonds of both the polymers.**

**Figure 11.** Change in isothermal **(a)** Overall crystallinity **(b)** Scaled crystallinity with number of Monte Carlo steps (*MCS*) for $\lambda$ **= 0, 1, 2, 3, 4, 5 and 6. The lines joining the points are meant only as a guide to the eye.**

**Figure 12. Development of initial crystallinity (up to 1000 *MCS*) for (a)** $\lambda = 0$ **and (b)** $\lambda = 2$ **with *MCS*, for A- and B-polymer. The lines joining the points are meant only as a guide to the eye.**

**Figure 13. Snapshots of semi-crystalline structures for** $\lambda = 1$ **(a) at** $U_p = 0.28$ **(during two-step isothermal cooling), (b) at** $U_p = 0.6$ **(during two-step isothermal cooling), (c) at** $U_p = 0.6$ **(one-step isothermal cooling). Blue lines represent crystalline bonds of A-polymer, magenta lines represent crystalline bonds of B-polymer and grey lines represent non-crystalline bonds of both the polymers.**

**Table 1**

| Component | Two-step cooling | | One-step cooling |
|---|---|---|---|
| | Up = 0.28 | Up = 0.6 | Up = 0.6 |
| A | 0.73 | 0.75 | 0.66 |
| B | 0.11 | 0.68 | 0.64 |

**Table 2**

| Component | Two-step cooling | | One-step cooling |
|---|---|---|---|
| | Up = 0.28 | Up = 0.6 | Up = 0.6 |
| A | 3.43 | 3.47 | 2.73 |
| B | 2.09 | 2.78 | 2.74 |

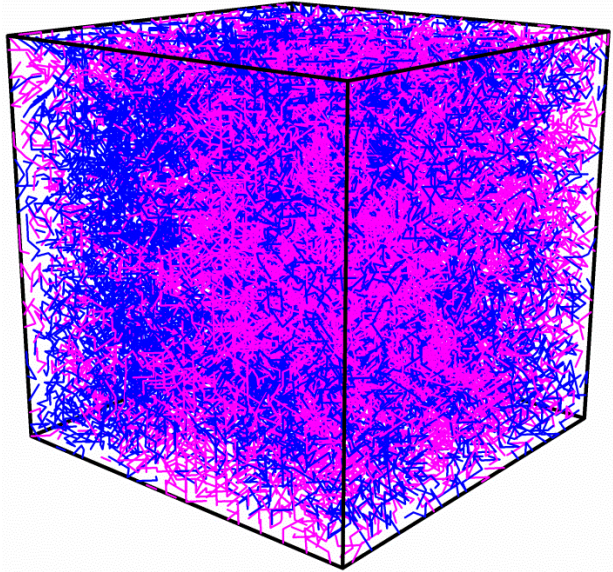

**Figure 1**

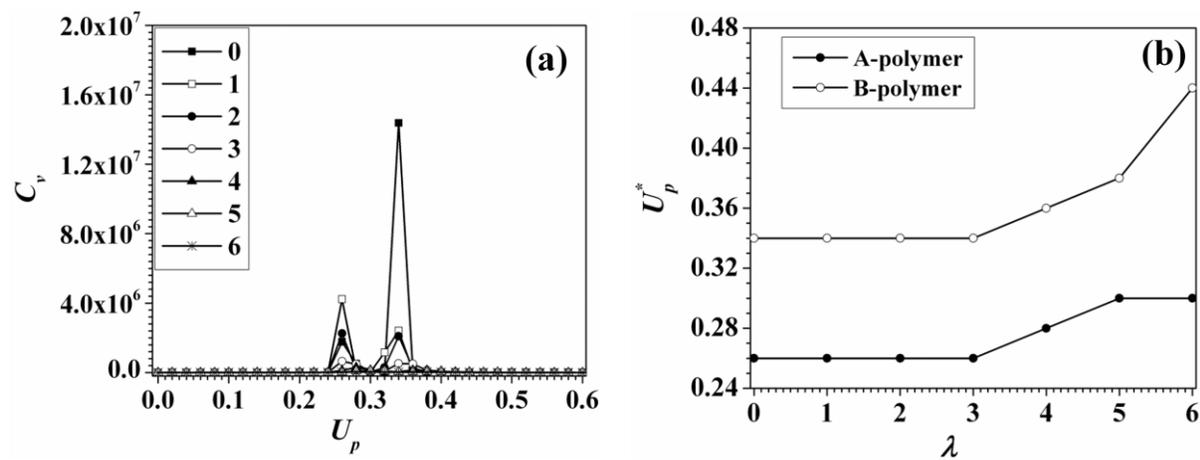

**Figure 2**

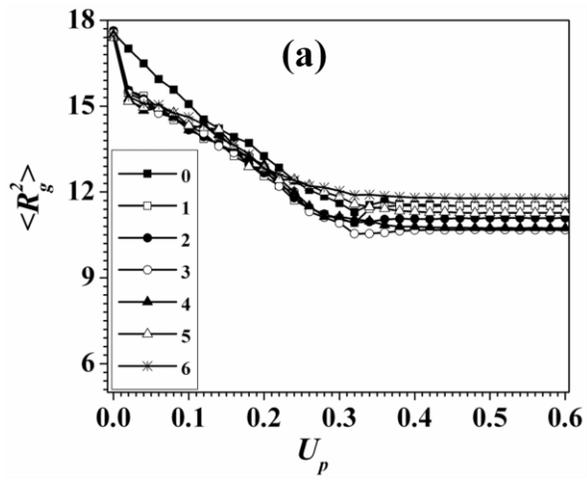 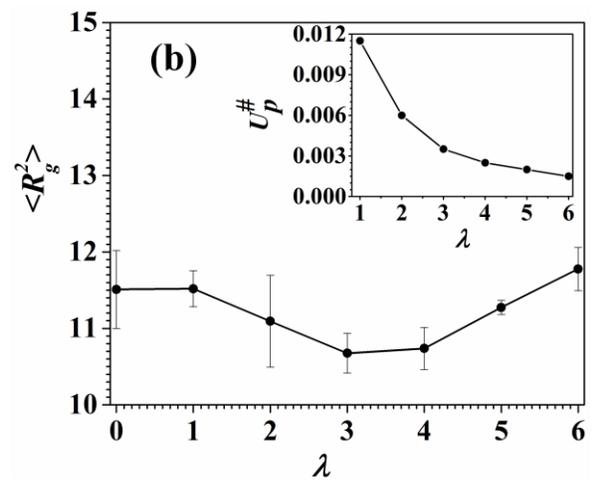

Figure 3

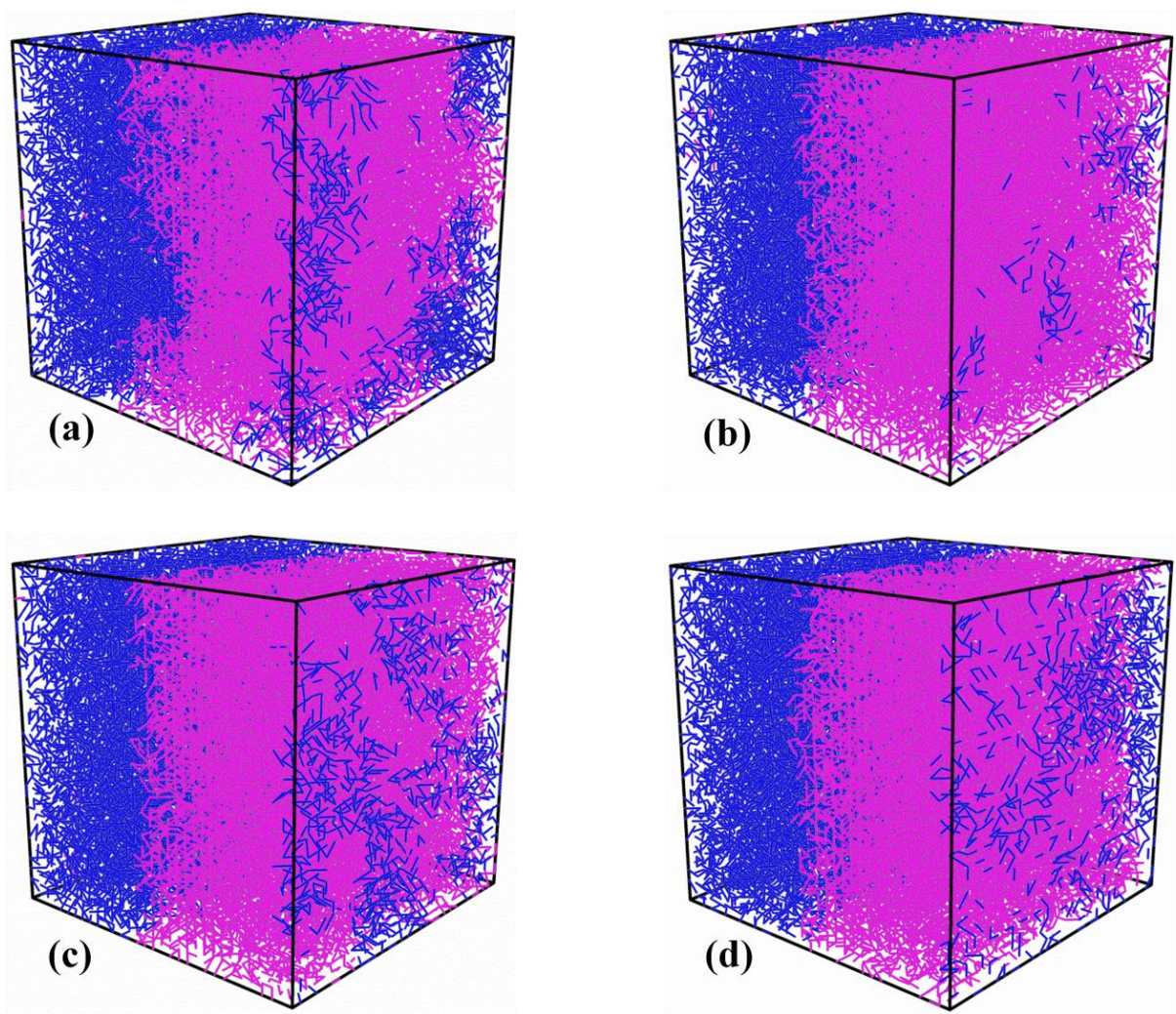

**Figure 4**

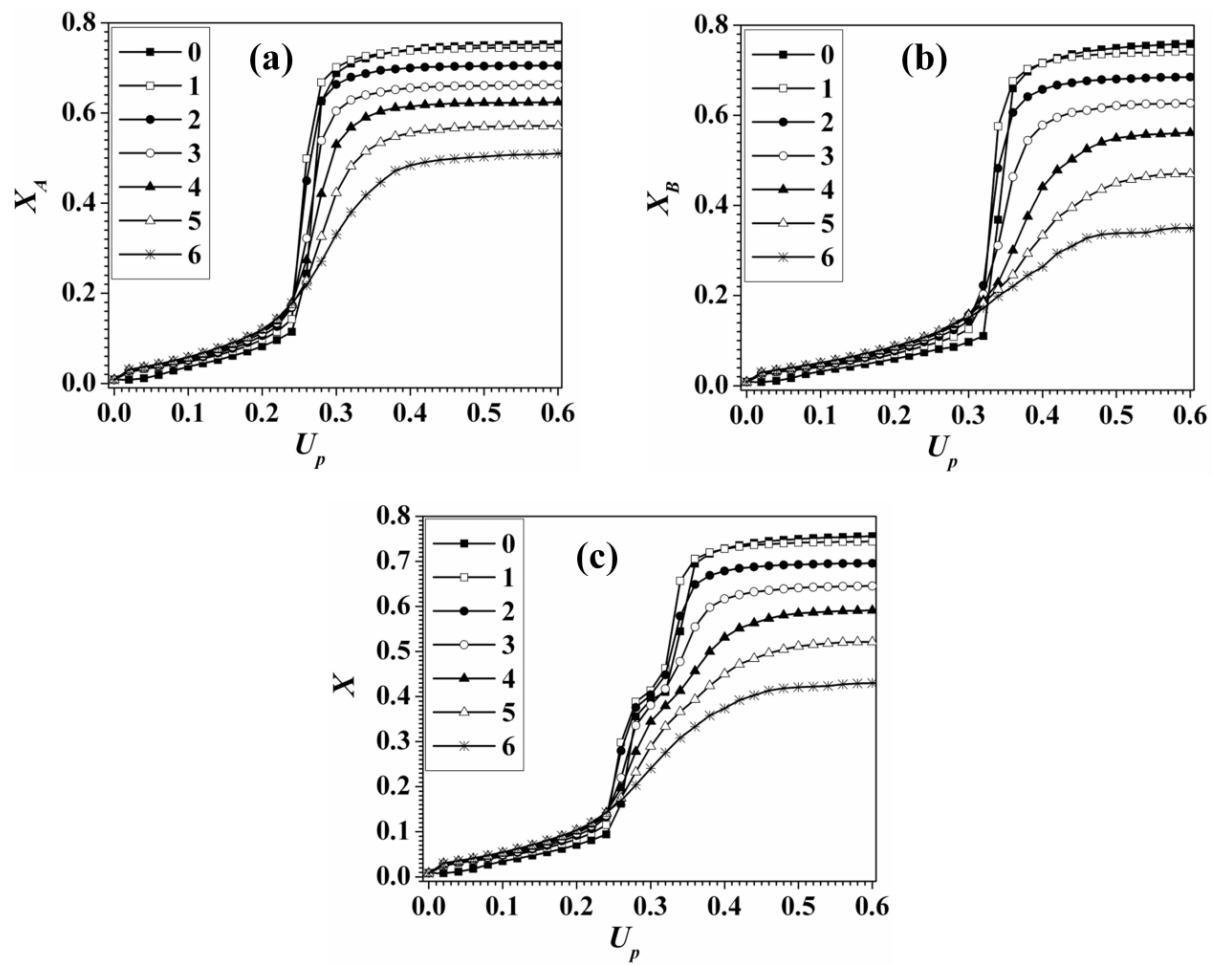

**Figure 5**

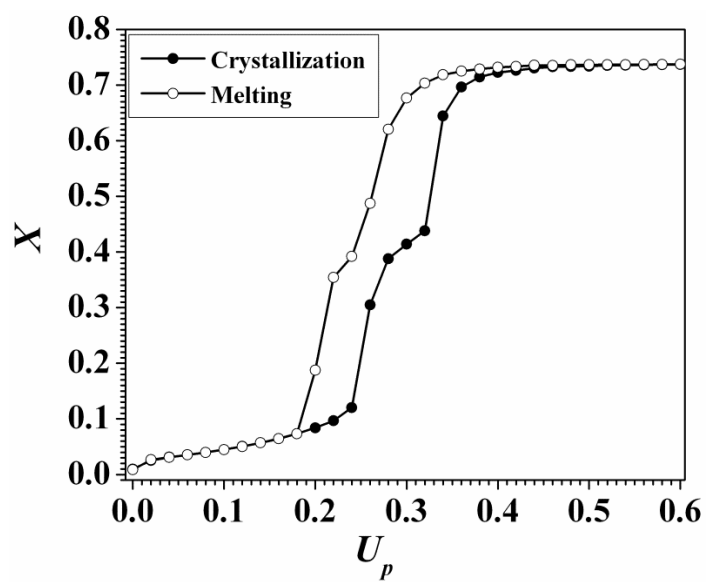

**Figure 6**

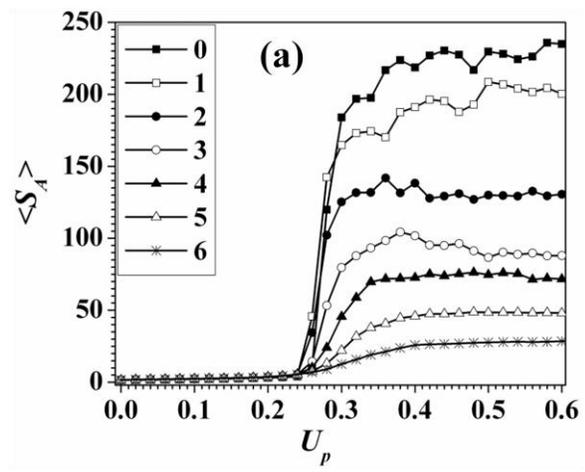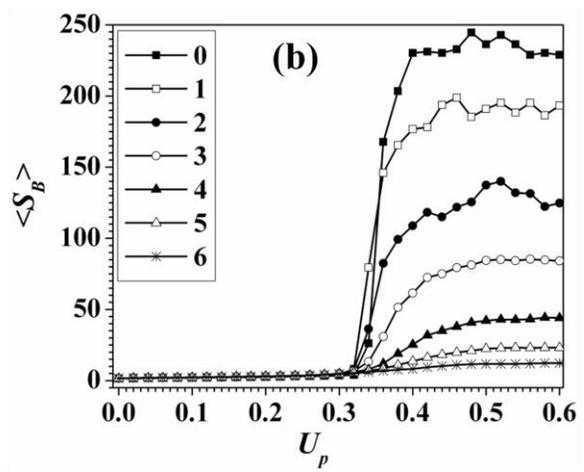

**Figure 7**

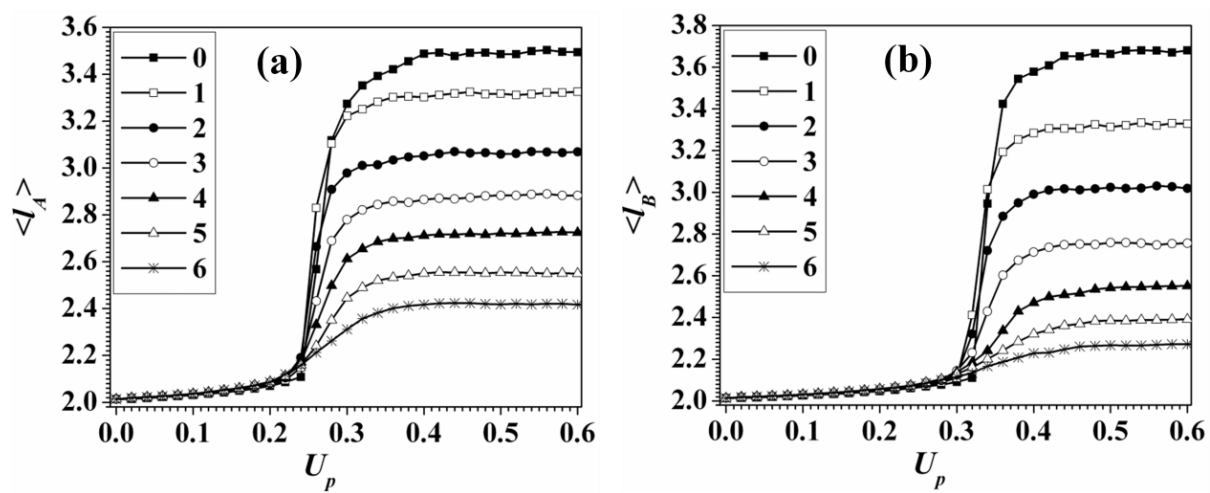

**Figure 8**

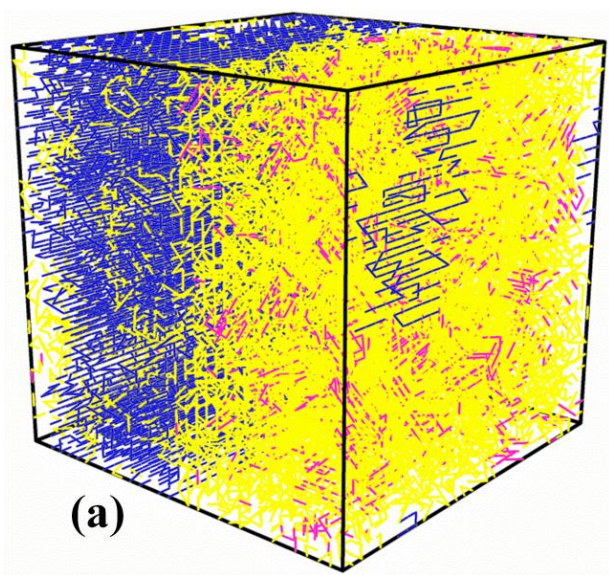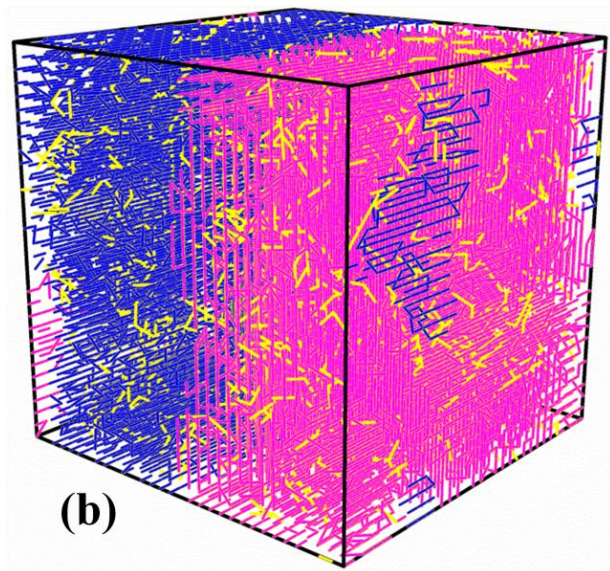

**(a)**  **(b)**

**Figure 9**

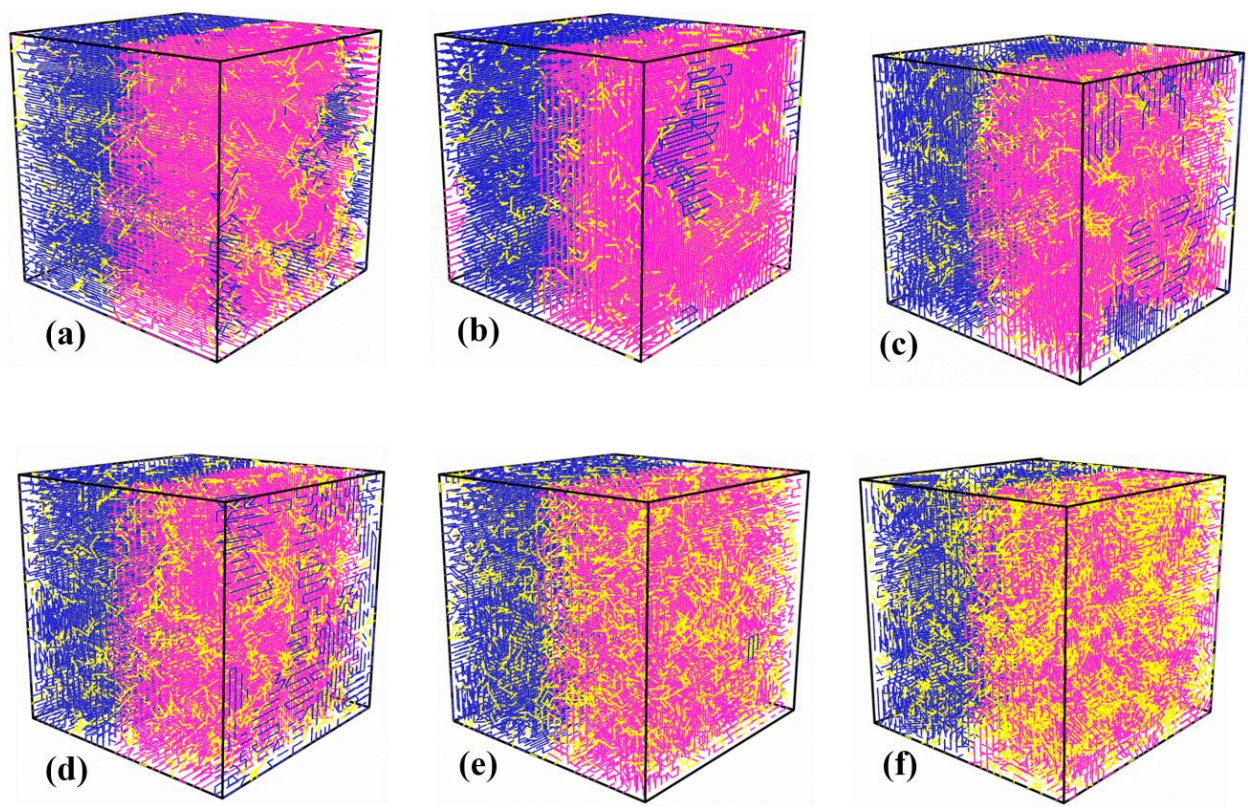

**Figure 10**

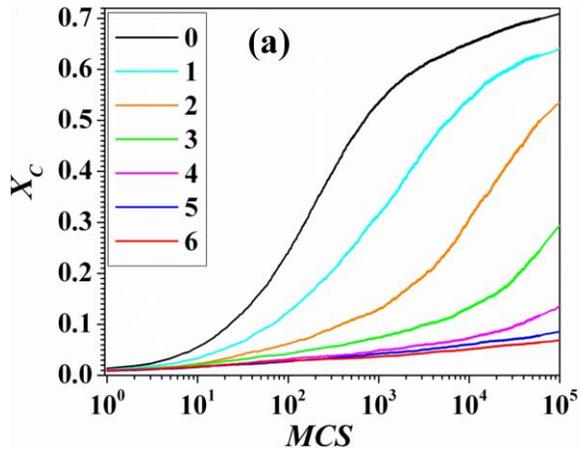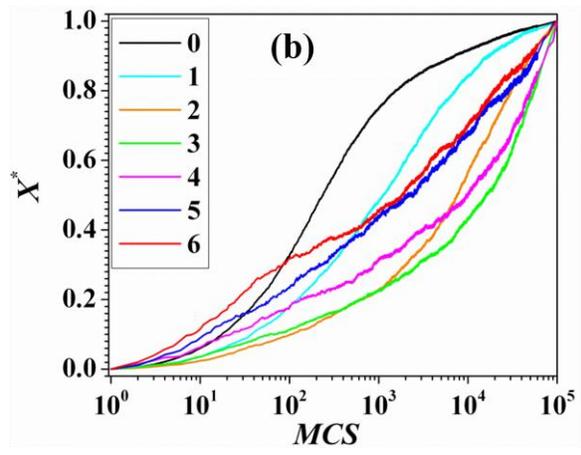

**Figure 11**

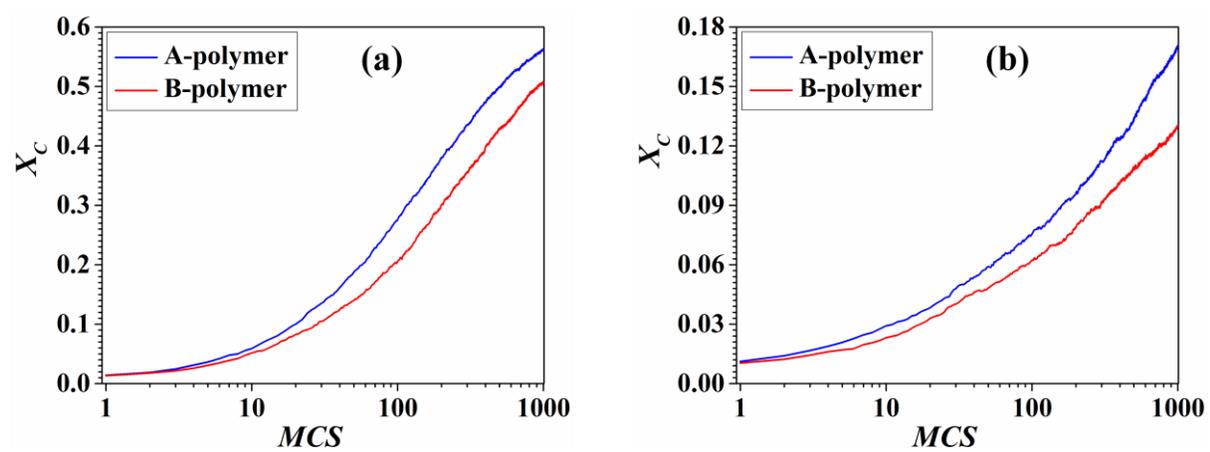

**Figure 12**

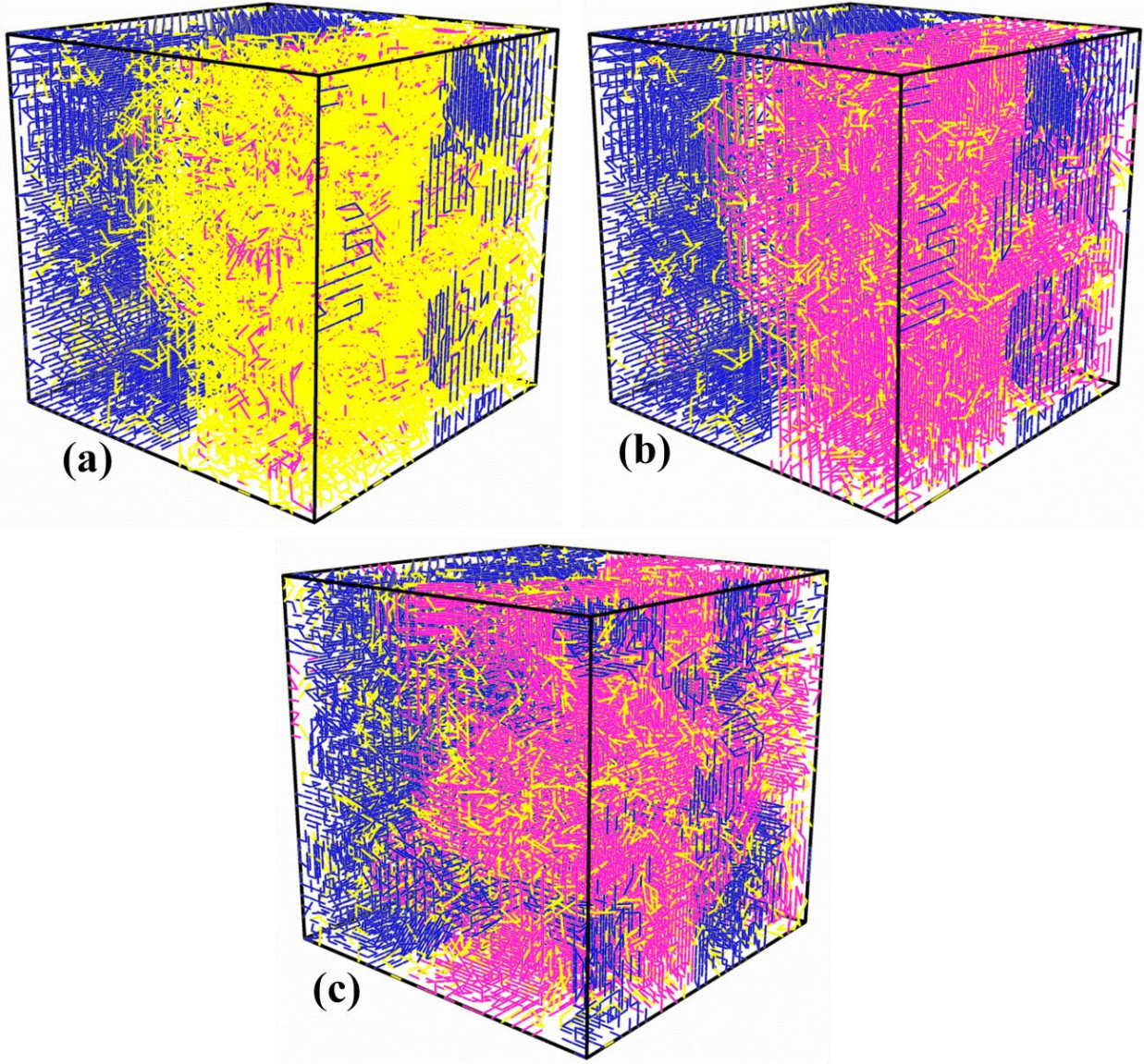

**Figure 13**